# Universal transfer and stacking technique of van der Waals heterostructures for spintronics

Yuan Cao,[1,§] Xinhe Wang,[1,§] Xiaoyang Lin,[1,2,3,†] Wei Yang,[1] Chen Lv,[1] Yuan Lu,[4] Youguang Zhang,[1] Weisheng Zhao[1,2,3,†]

[1] *Fert Beijing Research Institute, School of Microelectronics & Beijing Advanced Innovation Center for Big Data and Brain Computing (BDBC), Beihang University, Beijing 100191, China*
[2] *Beihang-Goertek Joint Microelectronics Institute, Qingdao Research Institute, Beihang University, Qingdao 266000, China*
[3] *Hefei Innovation Research Institute, Beihang University, Xinzhan Hi-tech District, Anhui 230013, China*
[4] *Institut Jean Lamour, UMR 7198, CNRS-Universite de Lorraine, F-54000 Nancy, France*
[§] Y. C. and X.H.W contributed equally to this work.
[†] *Address correspondence to XYLin@buaa.edu.cn (X.Y.L), weisheng.zhao@buaa.edu.cn (W.S.Z)*

The key to achieving high-quality van der Waals heterostructure devices made from various two-dimensional (2D) materials lies in the control over clean and flexible interfaces. However, existing transfer methods based on different mediators possess insufficiencies including the presence of residues, the unavailability of flexible interface engineering, and the selectivity towards materials and substrates since their adhesions differ considerably with the various preparation conditions, from chemical vapor deposition (CVD) growth to mechanical exfoliation. In this paper, we introduce a more universal method using a prefabricated polyvinyl alcohol (PVA) film to transfer and stack 2D materials, whether they are prepared by CVD or exfoliation. This peel-off and drop-off technique promises an ideal interface of the materials without introducing contamination. In addition, the method exhibits a micron-scale spatial transfer accuracy and meets special experimental conditions such as the preparation of twisted graphene and the 2D/metal heterostructure construction. We illustrate the superiority of this method with a $WSe_2$ vertical spin valve device, whose performance verifies the applicability and advantages of such a method for spintronics. Our PVA-assisted transfer process will promote the development of high-performance 2D-material-based devices.

## 1 Introduction

Since the observation of giant magnetoresistance (GMR) by Fert *et al.* and Grünberg *et al.*[1,2], spintronics, which utilizes the spin of electrons, has become a promising solution for next-generation data storage and information processing in the post-Moore's-law era[3,4]. After the first successful exfoliation of monolayer graphene in 2004[5], significant advances in the field of graphene spintronics, including efficient spin transport[6], spin relaxation[7,8], and spin logic devices[3,9] have been made. This progress has been followed by an impressive surge in the development of other two-dimensional (2D) materials, such as transition metal dichalcogenides (TMDs) [10–13]which further enrich 2D spintronics with features of gate tunability[14], spin-valley coupling[15], and even 2D ferromagnetism[16].

In parallel with the study of graphene-like 2D materials, van der Waals heterostructures fabricated by stacking 2D crystals on top of each other have also been gaining substantial attention[17–19]. These 2D heterostructures with fascinating properties could trigger new revolutions in innovative device and architecture designs for integrated circuit applications, such as GMR[20], and superconductivity[18]. In this sense, the development of a universal fabrication method that can achieve desired and clean interfaces between 2D materials, and fine contacts with electrodes has become a key issue with 2D electronics, especially 2D spintronics.

The existing fabrication methods of van der Waals heterostructures and devices[21–29] can be classified into wet and dry transfer methods. The polymethyl methacrylate (PMMA)-assisted[21] wet transfer method is most widely used, however, it engenders a major issue that the polymer contamination is difficult to remove after the transfer, which will degrade the performance of device. The dry transfer method represented by polydimethylsiloxane (PDMS) has been shown to reduce the contamination[30] and can achieve precise positioning transfer[31]. However, it is difficult for PDMS to pick up 2D atomic layers[32] to construct heterostructures, especially sandwich structures, and the stacking of more layers.

Here, we present a universal method utilizing a free-standing PVA film to realize 2D material transfer and fabrication of van der Waals heterostructures. Different from the two types of transfer methods represented by PMMA and PDMS, this method is clean, selective and accurate due to the water solubility of PVA and edge maneuverability during transfer. Special devices, such as magic-angle graphene and hybrid 2D/3D heterostructures, can also be constructed. The universality and reliability of the method are verified by further implementation of a vertical spintronic device. With the great potential of 2D van der Waals heterostructures fabrications, our method could promote the development of emerging spintronic devices.

## 2 Experimental

**Free-standing PVA-assisted Transfer and Stacking of 2D Materials.** As discussed above, the PMMA-assisted transfer method would cause polymer residuals and sample damage. The macromolecular transfer mediator typically requires removal in volatile toxic solvents such as acetone, which necessitates special handling and disposal precautions. However, PVA is a small-molecule and water-soluble mediator, which is environment-friendly and can maintain the sample's intrinsic properties during the removal process. Additionally, compared with the transfer method of spin coating polyvinyl pyrrolidone (PVP)[22] and PMMA[25], whereby the entire 2D material on the substrate is transferred at one time thus causing disturbance and inevitable waste, the selective transfer enabled by prefabricated PVA film is convenient and economizes materials. The sample could be a 2D material prepared by mechanical exfoliated or chemical vapor deposition (CVD) growth, even with the deposited metal on its surface. The transferred substrate can be a silicon wafer, a copper foil, quartz, or other substrates, as they can closely attach to the PVA.

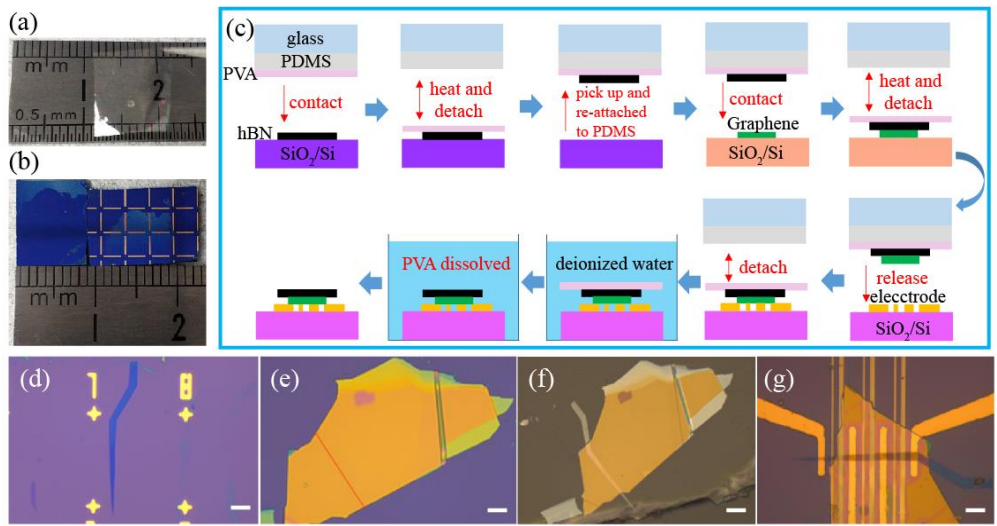

Figure. 1. (a) A free-standing PVA film handled with tweezers. (b) A chemical vapor deposition (CVD) grown $WSe_2$ film is transferred onto a marked $SiO_2$/Si substrate. (c) A schematic illustration of the PVA transfer processes. (d) & (e) Mechanically exfoliated graphene and h-BN. (f) An optical microscopy image of the PVA film/h-BN/graphene heterostructure. (g) An optical microscopy image of the as-fabricated device with the 2D heterostructure of h-BN/graphene placed on the pre-patterned electrodes. The scale bar of (d-g) is 10 μm.

The focus of the free-standing PVA assisted transfer is to prepare the PVA film of appropriate thickness (~50 μm, see Figure 1a). The much thinner one is easily curled which is not convenient for the transfer operation while the much thicker one will be difficult to be completely removed (see Methods for details of the preparation process of PVA film).

A variety of 2D materials have been transferred using the free-standing PVA thin film in our trials. The first is the overall transfer of the continuous monolayer $WSe_2$ on silicon substrate. The transfer process refers to the description in the next paragraph. The result is shown in Figure 1b, the whole film of CVD-grown $WSe_2$ is transferred at one time to the substrate with pre-patterned electrodes. This method also could be scaled up for the transfer of large-area 2D materials, provided the PVA film is large enough.

Further, the heterostructures of h-BN/graphene is prepared with our transfer method. The process is schematically illustrated in Figure 1c. After mechanical exfoliation, the h-BN and graphene crystals on silicon are selected for stacking under an optical microscope (Figure 1d and e). Then, take a glass slide with a piece of PDMS (10*10*2 mm) on it, to act as a releasable supporting platform for PVA. Next, a thin PVA film of the same size is taken and stuck onto the PDMS, whereby the PVA film could be easily bent and is adhered tightly to the PDMS due to the flexibility of PVA and the appreciable adhesion.

Under the microscope, with the help of a home-built transfer set with micromanipulation stage, the glass slide/PDMS/PVA is aligned and made to contact tightly with the as-selected h-BN flake. And heat the stage at 90°C for 2 minutes to guarantee the PVA film adheres to both the silicon wafer and h-BN flake firmly. The binding between PVA and PDMS is relatively weaker, so the PVA film adhered to the silicon wafer and can be separated from the PDMS slowly. On the other hand, the h-BN is adhered to the PVA stronger than to the silicon, so the PVA film carrying the h-BN flake can be peeled off slowly from the substrate with tweezers and reattached to PDMS to form a new stack of glass slide/PDMS/PVA/h-BN.

Similarly, the graphene can be picked up with the new stack, thanks to strong van der Waals interactions between both graphene/h-BN and graphene/PVA. The alignment between graphene and h-BN should be controlled carefully.

Finally, after transferring the stack of glass slide/PDMS/PVA/h-BN/graphene onto a silicon wafer with pre-patterned electrodes, and heating at 90°C for 2 minutes, the PVA/h-BN/graphene is firmly bonded to the top of the pre-patterned electrodes. By separating PDMS from PVA and dissolving the PVA with deionized water (at room temperature for 15 minutes), a h-BN/graphene heterostructure device (Figure 1g) is successfully fabricated.

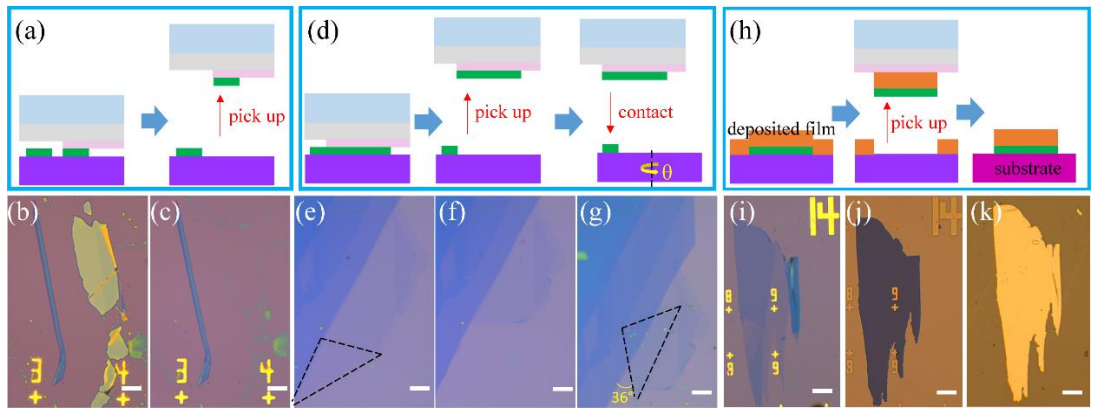

Figure. 2. Selective transfer and stacking with high spatial resolution. (a) A schematic illustration of a selective transfer process. (b) & (c) Exfoliated graphene fabricated by transferring the undesired part through the partial transfer method. The scale bar is 10μm. (d) Schematic illustration of the graphene with a twist that realized by partial transfers from monolayer graphene using the PVA film edge. (e-g) Optical images corresponding to the stacking process of (d). The scale bar is 10 μm. (h) A schematic illustration of a 2D material with a pre-deposited metal film transferred to another substrate. (i-k) Optical images corresponding to the transferring process in (h). Exfoliated graphene with deposited 30 nm Co and 5 nm Au films is transferred to a sputtered LSMO substrate. The scale bar is 15 μm.

## 3 Results and discussion

### 3.1 Selective transfer with spatial resolution of one micron

With the help of precision transfer technology and the prefilming process, our transfer method has a spatial resolution of approximate one micron and can be used to select 2D materials at a fixed position. Mechanically exfoliated 2D materials always appear in groups, adopting independent thin or single layer samples is challenging, which greatly complicates the fabrication process, especially the wire layout. The illustration of our selective transfer process with corresponding optical photographs are shown in Figure 2a-c. By controlling the contact area of the PVA film to match a specific area of 2D sample, the transfer of the selected sample can be achieved. The positioning accuracy can reach one micron. It should be noted that although the exposed PDMS also contacts other 2D materials elsewhere during the transfer process, the adhesion of PDMS is not as strong as that of the PVA. When separating the PVA with the selected sample from the substrate, other samples contacting the PDMS will not be taken away. Therefore, the transfer of the selected sample in a specific area can be realized without affecting surroundings, provided that the area of the PVA film is controlled.

### 3.2 Flexible stacking control of both angle and position

The interfacial engineering of 2D heterostructures, for example, by controlling the stacking order and angle, has endowed 2D materials/heterostructures with fascinating properties, such as anisotropic spin relaxation[8] and tunable spin absorption[9]. Now, utilizing a partial transfer of PVA film and a rotation of the substrate, we can easily realize these structures. Figure 2d describes the fabrication process of an angularly controlled stacked sample consisting of two rotationally aligned monolayer graphene flakes. The corresponding optical micrographs are shown in Figure 2e-g. The target graphene is partially picked up by controlling the contact area with the PVA film edge, subsequently, the remaining graphene is slightly rotated with the substrate by a rotator and adhered to the graphene that had just been picked up, forming a twist-angle bilayer graphene with controllable rotationally aligned. Compared with the requirement of making a hemispherical handle substrate[33] to transfer a sample with a specific area, this method only needs controlling the edge of the PVA to overlap with a fixed area, and greatly simplifies the process.

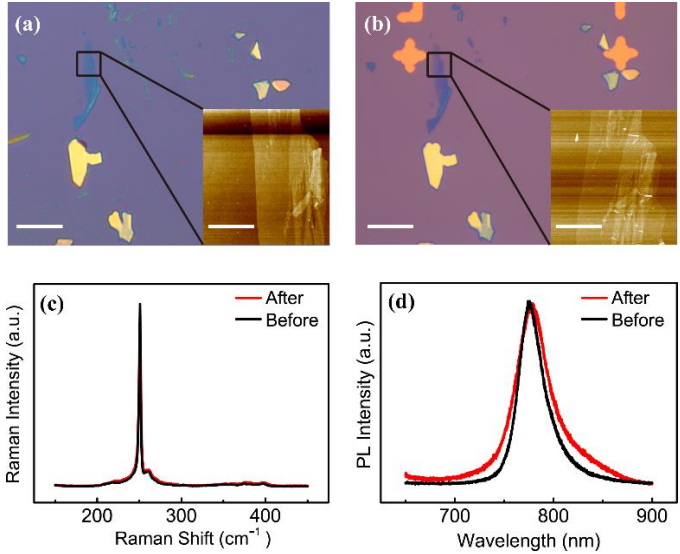

FIG. 3. (a) & (b) Optical images of exfoliated WSe$_2$ before and after transfer, and both regions are examined by AFM (insets), which indicates the cleanliness of the PVA transfer method. The scale bar is 10 μm in the optical images and 2 μm in the AFM images. (c) & (d) Typical Raman and PL spectra of single-layer CVD WSe$_2$ before (red line) and after (black line) the transfer.

### 3.3 Transfer of 2D materials with pre-deposited metal films

Contact engineering between 2D materials and bulk materials (such as metallic electrodes) has been a key issue of 2D electronics[34,35]. Conventional techniques are problematic in terms of impurities (optical lithography and electron beam lithography), which give rise to a large contact resistance. Several approaches[36,37] have been proposed but none has been fully developed. The PVA-assisted method may be able to improve this issue in vertical spin valves. Generally, the contamination of the contact interface comes from the lithography of electrode pattern before deposition. Figure 2h describes the transfer process of a heterostructure consisting a 2D material with a metal film deposited on it, which can avoid lithography contamination on the material's surface. The corresponding optical images are shown in Figure 2i-k. After the graphene with a deposited Co film (60 nm) is peeled off from the original SiO$_2$ substrate and transferred onto a LSMO magnetic substrate, a LSMO/graphene/Co spin valve is successfully prepared. Since the binding force between graphene and Co is larger than that between graphene and substrate[38], the deposited Co film and graphene are picked up together. The hybrid 3D/2D heterostructures prepared by this method may open a new way for future device fabrication.

### 3.4 Characterization and Device Performance

We use microscopic, atomic force microscopic (AFM), and spectroscopic approaches to evaluate the quality of this PVA transfer method. Taking the mechanically exfoliated WSe$_2$ as an example, we first mechanically exfoliated WSe$_2$ flakes onto a SiO$_2$/Si substrate and then selectively transferred a portion of the samples to another marked substrate. After the transfer, the target samples preserve their original state such as morphology and relative position (Figure 3a and b), and no breakages or wrinkles are found, which are further confirmed by AFM (inset of Figure 3a and b). The following are Raman and photoluminescence (PL) spectra characterization under 532nm excitation. The exfoliated WSe$_2$ clearly presents two Raman peaks in the range of 250 and 260 cm$^{-1}$ (Figure 3c), which correspond to the in-plane $E_{2g}^1$ mode and out-of-plane A$_{1g}$ mode of WSe$_2$. The Raman peaks are essentially coincident before and after the transfer process, indicating that no contamination is introduced to the transferred WSe$_2$. The PL peaks position before and after transfer display an identical band gap of 1.65 eV (Figure 3d), indicating that the band structure of the WSe$_2$ is not affected during the transfer. Though the full width at half maximum (FWHM) has slightly broadening from 29nm to 39nm, combined with substantially unchanged Raman and AFM data, this change can be caused by the interaction between the sample and different substrates[39].

The interface quality significantly affects the electronic structure and transport performance of 2D materials devices[40]. Especially for spintronic applications, the interface is a key issue[41]. To investigate the interface quality in our transfer method, we further fabricate a WSe$_2$ vertical spin valve in which bilayer WSe$_2$ is sandwiched between two ferromagnetic (FM) electrodes (NiFe and Co) and served as the nonmagnetic spacer layer[42]. As shown in Figure 4a, the resistance of junction is measured with 4-probe method. And the coercivities of two FM electrodes are designed to be different, so that the magnetization alignments of them can be made semi-parallel or semi-antiparallel by sweeping the in-plane magnetic field (H). The two resistance states corresponding to the two magnetization alignments give us the magnetoresistance (MR) of the spin valve.

The linear current-voltage (I–V) behaviors of the junction indicate the good Ohmic contacts of the NiFe/WSe$_2$/Co spin valve, and the resistance has positive correlation with temperature, exhibiting a metallic behavior similar to what has been reported for other TMD spin valve device[43–45]. A strong bonding is expected between WSe$_2$ and NiFe, yielding a strong wave-function overlap between W and NiFe states that results in a metallic junction. This expected bonding confirms that our method guarantees a 2D material to maintain a clean interface after transferred.

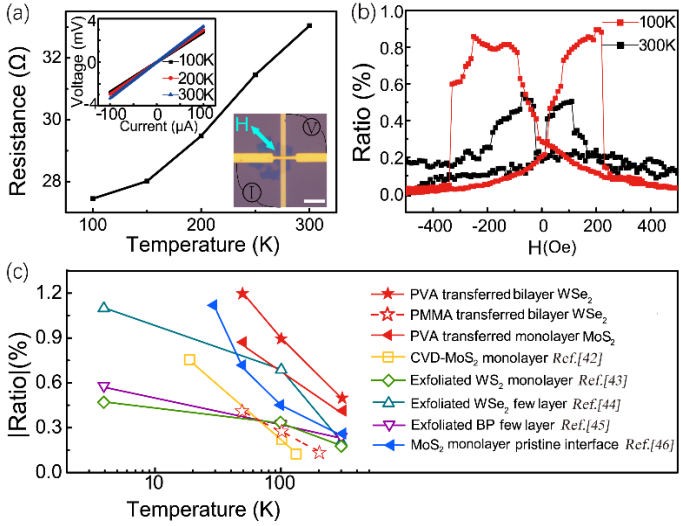

FIG. 4. (a) Resistance versus temperature of a WSe$_2$ spin valve. The insets present the I-V curves from a four-probe measurement and the corresponding device diagram. The scale bar is 10 μm. (b) The magnetoresistance (MR) ratio versus temperature of a WSe$_2$ spin valve at 100 K and 300 K. (c) The distribution of absolute MR ratios of 2D vertical spin valves prepared by different transfer methods at various temperatures.

A typical MR curve of our Co/WSe$_2$/NiFe spin valves at 300 K and 100 K are shown in Figure 4b. There are two states, parallel alignment of the magnetization for low resistance (R$_P$) and antiparallel alignment for high resistance (R$_{AP}$). The MR ratio is then defined as [R$_{AP}$ - R$_P$]/R$_P$, which is 1.2% at 50 K and 0.5% at 300 K, as shown in Figure 4c. For evaluating this result, we compare the MR ratios at various temperatures that obtained in different vertical spin valves of 2D materials, and their structures are uniformly FM/2D/FM (FM=Co or NiFe. Those modified structures are excluded, such as adding a tunneling layer for enhancing the MR ratio). Our spin valves of PVA transferred bilayer WSe$_2$ show a relatively highest range of MR ratios (solid pentagram). For comparison, we also prepare the same device with conventional PMMA wet transfer (open pentagram), whose range of MR ratios is within the reported results from various 2D spin valves that are all based on conventional wet transfer (open symbols)[43–46]. By contrast, these results from conventional wet transfer are all significantly lower than the results from PVA transferred bilayer WSe$_2$.

Additionally, the MR range labeled by blue solid triangles in Figure 4c is from the monolayer-MoS$_2$ spin valve with pristine interfaces, as reported in [47], in which the contacts are formed by directly depositing metal on the pristine exfoliated 2D material, through a hole in the bottom of the substrate. For making a direct comparison, we prepare the same device by PVA transferring, whose MR range is shown with red solid triangles, which is reached to the level of the above devices with pristine interface. The results indicate that in our transfer process, after dissolving the PVA, the surface of the 2D material is ultraclean and approaches intrinsic quality.

## 4 Conclusions

In conclusion, we develop a method for transferring and stacking 2D-material van der Waals heterostructures via a free-standing and water-soluble PVA film. The proposed method can selectively control the angle and position of the stacking sample, which has a one-micron spatial resolution. This precision meets the requirements for the interface and contact engineering of 2D heterostructures. AFM, Raman and PL characterizations show the cleanliness of the surface after the transfer. Furthermore, we construct the vertical spin valves based on 2D materials, which show superior performances guaranteed by our transfer method. Such a universal transfer method can promote the development of future spintronic devices.

## 5 Methods

**PVA film preparation process**

Dissolve the PVA powder (Alfa Aesar, 98-99% hydrolyzed, high molecular weight) in deionized water with a mass ratio of 4%~6%. Decant the solution into a Petri dish with the liquid height being about 1 mm. Place the dish in the fume hood to accelerate the evaporation of the solvent. The PVA film could be formed after a few hours. Subsequently, the as-prepared PVA could be cut and peeled off as a free-standing film.

## Acknowledgements

This work was supported by the National Natural Science Foundation of China (No. 51602013, 11804016, 61627813, and 11904014), Young Elite Scientists Sponsorship Program by China Association for Science and Technology (CAST) (No. 2018QNRC001), the International Collaboration 111 Project (No. B16001), the China Postdoctoral Science Foundation (No. 2018M641152, BX20180022), the Fundamental Research Funds for the Central Universities of China, and the Beijing Advanced Innovation Centre for Big Data and Brain Computing (BDBC).